\documentclass[fleqn,twoside]{article}
\usepackage{espcrc2}
\usepackage{epsfig}

\usepackage{graphicx}
\usepackage[figuresright]{rotating}

\newcommand{\AmS}{{\protect\the\textfont2
  A\kern-.1667em\lower.5ex\hbox{M}\kern-.125emS}}

\title{Top quark pair production and decay at hadron colliders: Predictions
at NLO QCD including spin correlations\thanks{Report 
DESY 02-144, PITHA 02/15, TTP02-24}}

\author{W. Bernreuther\address[RWTH]{Institut f. Theoretische Physik,  
        RWTH Aachen, 52056 Aachen, Germany}\thanks{supported by BMBF contract 05 HT1 PAA 4},
        A. Brandenburg\address{DESY-Theorie, 22603 Hamburg, Germany}\thanks{Speaker at the conference. 
         Supported by a Heisenberg fellowship of D.F.G.}, 
Z.G. Si\addressmark[RWTH]$^{*}$,
        P. Uwer\address{Institut f. Theoretische Teilchenphysik, 
        Universit\"at Karlsruhe, 76128 Karlsruhe, Germany}
      }
       
\begin{document}
\begin{abstract} 
We present results at NLO QCD for hadronic production and decay
of top quark pairs, taking into account $t\bar{t}$ spin correlations.
\vspace{1pc}
\end{abstract}

\maketitle

The interactions of the top quark have not been precisely tested so far. 
The expected data samples of Run II of the Tevatron collider
(eventually about $10^4$ $t\bar{t}$ pairs per annum) and the LHC
($\sim \ 10^7$ $t\bar{t}$ pairs p.a.) will make possible detailed 
investigations of top quark production and decay. In particular,
one may search for
possible deviations from the QCD production mechanism and the 
V-A structure of the top decay by analysing
energy and angular distributions of the $t\bar{t}$ decay products.
Such studies will be feasible since top quarks do not hadronize due
to their very short lifetime, and perturbation theory gives a reliable
description of top quark production and decay. 
To search for new physics effects, 
precise theoretical 
predictions for top quark pair production and decay at hadron colliders
within the Standard Model are needed.
We consider here in particular the following processes:
\begin{equation}
h_1h_2\to t\bar{t}+X\to
\left\{\begin{array}{c}
\ell^+\ell '^{-} + X\\ 
\ell^+j_{\bar{t}} + X\\ 
\ell^-j_{t} + X\\
j_t j_{\bar{t}}+X,
 \end{array} \right. 
\label{reac1}
\end{equation}
where $h_{1,2}=p,\bar{p}$; $\ell=e,\mu,\tau$, and 
$j_t\ (j_{\bar{t}})$ denote  
jets originating from hadronic $t$ ($\bar{t}$) decays. 
An observable that is very sensitive to the dynamics of top quark production
and decay in the above reactions is the double differential angular
distribution of the top decay products, e.g. for the dilepton channel:
\begin{equation}\label{dist}
\frac{1}{\sigma}\frac{d^2\sigma}{d\cos\theta_+d\cos\theta_-}
=\frac{1}{4}(1-C\cos\theta_+\cos\theta_-).
\end{equation}
In (\ref{dist}), $\theta_\pm$ are the angles between the $\ell^{\pm}$
direction of flight in the $t(\bar{t})$ rest frame with respect to
arbitrary axes $\hat{\bf a}$ ($\hat{\bf b}$) which will be specified below. 
(Terms linear in $\cos\theta_\pm$ are absent in (\ref{dist})
for our choices of  $\hat{\bf a}$, $\hat{\bf b}$ due to parity invariance
of QCD.)  

The calculation of the distribution (\ref{dist}) at next-to-leading order 
(NLO) QCD simplifies 
enormously in the leading pole
approximation (LPA), which amounts to expanding the full
amplitudes for (\ref{reac1}) around the complex poles of the $t$ and
$\bar{t}$ propagators. Only the leading pole terms are kept in this 
expansion, i.e. one neglects terms of order $\Gamma_t/m_t\approx 1\%$.
Within the LPA, the radiative corrections can be classified into
factorizable and non-factorizable \cite{beenakker} contributions.
Here we consider only the factorizable corrections. 
In this case the coefficient in the distribution (\ref{dist}) factorizes:
\begin{equation}\label{c}
C = \kappa_+\kappa_-D.
\end{equation}
In (\ref{c}), $D$ is the $t\bar{t}$ double spin asymmetry
\begin{equation}
\label{d}
D=\frac{N(\uparrow\uparrow)+N(\downarrow\downarrow)-N(\uparrow\downarrow)
-N(\downarrow\uparrow)}{N(\uparrow\uparrow)
+N(\downarrow\downarrow)+N(\uparrow\downarrow)+N(\downarrow\uparrow)},
\end{equation}
where $N(\uparrow\uparrow)$ denotes the number of $t\bar{t}$ pairs
with $t$ ($\bar{t}$) spin parallel to the reference axis $\hat{\bf a}$ 
($\hat{\bf b}$) etc. The directions $\hat{\bf a}$ and $\hat{\bf b}$  
can thus be identified with the
spin quantization axes for $t$ and $\bar{t}$, and $D$ reflects the
strength of the correlation between the $t$ and $\bar{t}$ spins 
for the chosen axes.
Further, $\kappa_\pm$ is the spin analysing power
of the charged lepton in the decay 
$t(\bar{t})\to b(\bar{b})\ell^\pm\nu (\bar{\nu})$ 
defined by the decay distribution
\begin{equation}\label{decay}
\frac{1}{\Gamma}\frac{d\Gamma}{d\cos\vartheta_{\pm}}=\frac{1\pm\kappa_\pm
\cos\vartheta_\pm}{2},
\end{equation}
where $\vartheta_\pm$ is the angle between the $t$ $(\bar{t})$ spin
and the $\ell^\pm$ direction of flight.
From \cite{cjk} we obtain
\begin{eqnarray} 
\kappa_+=\kappa_-=1-0.015\alpha_s,
\end{eqnarray}
which means that the charged lepton serves as a perfect analyser 
of the top quark spin. 
For hadronic decays $t\to bq\bar{q}'$ one has a decay distribution analogous
to (\ref{decay}), and the spin analysing power of jets can be defined.
To order $\alpha_s$ \cite{bsu} (and using $\alpha_s(m_t)=0.108$), 
\begin{eqnarray} \label{kappa}
\kappa_b=-0.408\times(1-0.340\alpha_s)=-0.393,\\
\kappa_j=+0.510\times(1-0.654\alpha_s)=+0.474,
\end{eqnarray}   
where $\kappa_j$ is the analysing power of the least energetic non-b-quark
jet. 
The results for $C$ given below are for the dilepton channel. The corresponding
results for the single lepton channel are obtained by replacing
$\kappa_+$ (or  $\kappa_-$) in Eq.~(\ref{c}) by
$\kappa_b$ or $\kappa_j$. The loss of analysing power 
using hadronic final states is 
overcompensated by the gain
in statistics and in efficiency to reconstruct the $t$ ($\bar{t}$) rest frames.

To compute $D$  at NLO QCD, the differential
cross sections for the following parton processes are needed to order
$\alpha_s^3$, 
where the full
information on the $t$ and $\bar{t}$ spins has to be kept:  
\begin{eqnarray}\label{parton} 
q\bar{q}\to t\bar{t},\ t\bar{t}g;\ 
gg\to t\bar{t},\  t\bar{t}g;\  
q(\bar{q})g\to t\bar{t}q(\bar{q}).
\end{eqnarray}
Results  at NLO QCD for the $\overline{{\mbox{MS}}}$ 
subtracted parton cross sections
$\hat{\sigma}$ for the above processes
with $t\bar{t}$ spins summed over have been obtained in \cite{nason}, 
\cite{been}, while 
\begin{equation}
\hat{\sigma}D= \hat{\sigma}(\uparrow\uparrow)+
\hat{\sigma}(\downarrow\downarrow)-\hat{\sigma}(\uparrow\downarrow)
-\hat{\sigma}(\downarrow\uparrow)
\end{equation}
has been computed for 
different spin quantization axes in \cite{us1}, \cite{us2}.
This combination of spin-dependent cross sections can be written as
follows
\begin{eqnarray}
\hat{\sigma}D=\frac{\alpha_s^2}{m_t^2}\left[
g^{(0)}(\eta) +4\pi\alpha_s {\cal{G}}^{(1)}\right],
\end{eqnarray}
with
\begin{eqnarray}
{\cal{G}}^{(1)}=g^{(1)}(\eta)+
\tilde{g}^{(1)}(\eta)\ln\left(\frac{\mu^2}{m_t^2}\right),
\end{eqnarray}
where $\eta=\hat{s}/(4m_t^2)-1$, and we use a common scale $\mu$ for 
the renormalization and factorization scale. As an example, Fig. 1 shows, 
for the parton process $gg\to t\bar{t}(g)$,   
the three functions that determine 
$\hat{\sigma}D$
for the helicity basis, i.e. $\hat{\bf a}$ ($\hat{\bf b}$) is chosen to be 
the $t$ ($\bar{t}$) direction of flight.
\begin{figure}[t]
\unitlength1.0cm
\begin{center}
\vspace{-0.6cm}
\begin{picture}(8,8)
\put(0,0){\psfig{figure=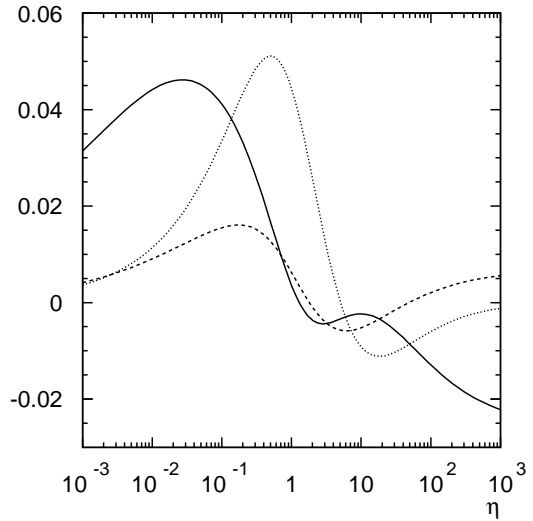,width=8cm,height=8cm}}
\end{picture}
\vspace{-1cm}
\caption{Dimensionless scaling functions $g^{(0)}(\eta)$
(full), $g^{(1)}(\eta)$ (dotted), and
${\tilde g}^{(1)}(\eta)$ (dashed) that determine 
$\hat{\sigma}D$ in the helicity basis 
for the $gg$ initial state.}\label{fig:siggg}
\vspace{-0.75cm}
\end{center}
\end{figure}
Apart from the helicity basis we also consider the beam basis, where
$\hat{\bf a}$ and $\hat{\bf b}$ are equal to the direction of one of the
hadron beams in the laboratory frame, and the so-called off-diagonal
basis \cite{off}, which is defined by the requirement that 
$\hat{\sigma}(\uparrow\downarrow)=\hat{\sigma}(\downarrow\uparrow)=0$
for the process $q\bar{q}\to t\bar{t}$ at tree level.
\begin{table}[h!]
\vspace{-0.3cm}
\caption{Coefficient $C$ defined in Eq.~(\ref{dist}) 
to leading and next-to-leading
order in $\alpha_s$ for the parton distribution functions of \cite{CTEQ5}
and $\mu=m_t=175$ GeV.}
\label{table:1}
\renewcommand{\arraystretch}{1.2} 
\begin{center}
\begin{tabular}{ccccc}
\hline
 &\multicolumn{2}{c}{$p\bar p$, $\sqrt{s}=2$~TeV }
&\multicolumn{2}{c}{$pp$, $\sqrt{s}=14$~TeV} \\
& LO & NLO & LO & NLO\\  \hline
$C_{\rm hel.}$ & $-0.456$& $-0.389$ & $\hphantom{-}0.305$  &
$\hphantom{-}0.311$\\
$C_{\rm beam}$ & $\hphantom{-}0.910$&  $\hphantom{-}0.806$ &
$-0.005$ & $-0.072$\\
$C_{\rm off.}$ & $\hphantom{-}0.918$ & $\hphantom{-}0.813$ &
$-0.027$ & $-0.089$\\ \hline
\end{tabular}
\vspace{-0.5cm}
\end{center}
\end{table}
\begin{table}[h!]
\vspace{-0.3cm}
\caption{Upper part: Dependence on $\mu$ of the correlation coefficients
$C$, computed at NLO with the PDF  of ref.\protect\cite{CTEQ5}. 
Lower part:
Correlation coefficients $C$ 
at NLO for $\mu=m_t$ and different sets
of parton distribution functions:  GRV98\protect\cite{Gluck:1998xa}, 
CTEQ5\protect\cite{CTEQ5}, and MRST98 (c-g)\protect\cite{MRST98}.} 
\par
\begin{center}
\renewcommand{\arraystretch}{1.2}
{\begin{tabular}{ccccc} \hline
& \multicolumn{3}{c}{Tevatron}
& LHC \\
 $\mu$   & $C_{\rm hel.}$ &  $C_{\rm beam}$ &
 $C_{\rm off.}$ &  $C_{\rm hel.}$ \\ \hline
$m_t/2$ & $-0.364$   & 0.774    & 0.779  &   0.278
\\
$m_t$   & $-0.389$   & 0.806  & 0.813 &  0.311
\\
$2m_t$ & $-0.407$  & 0.829 & 0.836  &  0.331
\\ \hline \hline
 PDF   & $C_{\rm hel.}$ &  $C_{\rm beam}$ &
 $C_{\rm off.}$ &  $C_{\rm hel.}$ \\ \hline
GRV98 & $-0.325$   & 0.734    & 0.739  &  0.332
\\
CTEQ5   & $-0.389$   & 0.806  & 0.813 &  0.311
\\
MRST98 & $-0.417$  & 0.838 & 0.846  &  0.315
\\ \hline
\end{tabular}}\label{tab:mudep}
\vspace{-0.75cm}
\end{center}
\end{table}
 
Table 1 lists our results \cite{us3} for the coefficient $C$ in the double 
differential distribution (\ref{dist}) for these 
different choices of spin quantization axes, 
using the parton distribution functions (PDFs) of \cite{CTEQ5}. 
One sees that at the Tevatron, the spin correlations are largest in the
beam and off-diagonal basis. The QCD corrections reduce the LO results by 
about 10\%. For the LHC, the best choice of the above three bases 
is the helicity basis, and 
the QCD corrections are small in that case.

Table 2 shows the dependence of the NLO results for $C$ on the scale
$\mu$ (upper part) and on the choice of the PDFs (lower part).
At the Tevatron the spread of results for different PDFs is larger
than the scale uncertainty. This rather strong dependence on the PDFs is 
largely due to the fact that the contributions from $q\bar{q}$ and $gg$ initial
states contribute to $C$ with opposite signs. This may offer the possibility
to constrain the quark and gluon content of the proton by a precise 
measurement of the double angular distribution (\ref{dist}).

In all results above we used $m_t=175$ GeV. A variation of 
$m_t$ from $170$ to $180$ GeV changes the results for the 
Tevatron, again for $\mu=m_t$ and PDFs of \cite{CTEQ5} as follows: 
$C_{\rm hel.}$ varies from $-0.378$ to $-0.397$, $C_{\rm beam}$
from $0.790$ to $0.817$, and  $ C_{\rm off.}$ from $0.797$ to
$0.822$. For the LHC,  $C_{\rm hel.}$ changes by less than a percent.  

The results above have been obtained without imposing any kinematic cuts.
Standard cuts on the top quark transverse momentum and rapidity only have
a small effect on $C$: For the Tevatron, demanding 
$|{\bf k}_{t,\bar{t}}^T|>15$ GeV and $|r_{t,\bar{t}}|<2$ leads to 
the following results: 
$C_{\rm hel.}=-0.386,\ C_{\rm beam}=0.815,\ C_{\rm off.}=0.823$.
For the LHC, when imposing the cuts $|{\bf k}_{t,\bar{t}}^T|>20$ GeV 
and $|r_{t,\bar{t}}|<3$, we find $C_{\rm hel.}=0.295$.

In summary, $t\bar{t}$ spin correlations are large effects that can
be studied at the Tevatron and the LHC by measuring double angular 
distributions both in the dilepton and single lepton decay channels. The
QCD corrections to these distributions have been computed and are
under control. Spin correlations are suited to analyse in detail top quark
interactions, search for new effects, and may help to constrain the
parton content of the proton.

\end{document}